\documentclass[%
 reprint,
 amsmath,amssymb,
 aps,
]{revtex4-1}

\usepackage{graphicx}
\usepackage{dcolumn}
\usepackage{bm}
\usepackage{threeparttable}
\usepackage{multirow}
\usepackage[T1]{fontenc} 

\begin{document}

\preprint{AP1roG-LCCSD}

\title{Linearized Coupled Cluster Correction on the Antisymmetric Product of 1 reference orbital Geminals}

\author{Katharina Boguslawski}
\email{katharina.boguslawski@gmail.com}
\affiliation{
Institute of Physics, Faculty of Physics, Astronomy and Informatics, Nicolaus Copernicus University, Grudzi\k{a}dzka 5, 87-100 Toru\'{n}, Poland
}
\author{Paul W.~Ayers}
\affiliation{
Department of Chemistry and Chemical Biology, McMaster University, Hamilton, 1280 Main Street West, L8S 4M1, Canada
}

\date{\today}

\begin{abstract}
We present a Linearized Coupled Cluster (LCC) correction based on an Antisymmetric Product of 1 reference orbital Geminals (AP1roG) reference state. In our LCC ansatz, the cluster operator is restricted to double and to single and double excitations as in standard single-reference CC theory. The performance of the AP1roG-LCC models is tested for the dissociation of diatomic molecules (C$_2$ and F$_2$), spectroscopic constants of the uranyl cation (UO$_2^{2+}$), and the symmetric dissociation of the H$_{50}$ hydrogen chain. Our study indicates that an LCC correction based on an AP1roG reference function is more robust and reliable than corrections based on perturbation theory, yielding spectroscopic constants that are in very good agreement with theoretical reference data.
\end{abstract}

\maketitle


\section{Introduction}
In the past decade, quantum-chemical studies prove helpful in understanding and predicting chemical phenomena of small and medium-sized molecules~\cite{Roos_U2,jankowski2012theory}. 
Routine applications to larger molecular systems are, however, hampered as conventional quantum chemistry methods are either too expensive or too approximate to guarantee reliable results. These problems are particularly severe for molecules where strong electron correlation becomes important. Examples for strongly-correlated systems are, for instance, radicals, transition metals, and actinide compounds. Furthermore, well-established approaches for strong correlation scale poorly, often factorially, with system size.
These drawbacks motivate the development of new, unconventional electron correlation methods for strongly-correlated many-body systems that represent computationally cheaper alternatives to standard methods.

One active field of research for strongly-correlated systems focuses on the development of wavefunction methods that use two-electron functions (geminals) to model the correlated motion of electrons~\cite{Rassolov_2002,Surjan_2012,Johnson_2013,Ellis_2013,piris2014perspective}.
The most popular geminal-based approaches are the Antisymmetric Product of Strongly orthogonal Geminals~\cite{Hurley_1953,Parr_1956,Parks_1958,Kutzelnigg_1964,Kutzelnigg_1965,Rassolov_2002,Pernal2013,jeszenszki2015local} (APSG), the Antisymmetrized Geminal Power~\cite{Coleman_1965,Coleman_1997,neuscamman2012size,neuscamman2013jastrow} (which is a special case of projected Hartree--Fock--Bogoliubov~\cite{PHF}), the Antisymmetric Product of Interacting Geminals~\cite{Bratoz1965,Silver_1969,Silver_1970,APIG-1,APIG-2,Surjan-bond-1984,Surjan-bond-1985,Surjan-bond-1994,Surjan-bond-1995,Surjan-bond-2000,Surjan_1999,Rosta2002,Surjan_2012} (APIG), Generalized Valence Bond~\cite{Hurley_1953,GVB1,GVB2,VB-CC,benzene-Gordon} (GVB), and the Antisymmetric Product of 1-reference orbital Geminals (AP1roG)~\cite{Limacher_2013,OO-AP1roG,Tamar-pCC}. Specifically, AP1roG provides a very good approximation to the doubly occupied configuration interaction~\cite{DOCI} (DOCI) wavefunction, at mean-field computational cost. The AP1roG geminal creation operator reads
\begin{equation}
\phi_i^{\dagger}=\text{a}_i^{\dagger} \text{a}_{\bar{i}}^{\dagger}+\sum_a \text{c}_i^a \text{a}_a^{\dagger} \text{a}_{\bar{a}}^{\dagger},
\end{equation}
where $a_{p}^{\dagger}$, $a_{\bar{p}}^{\dagger}$ are the standard fermionic creation operators for spin-up ($p$) and spin-down electrons ($\bar{p}$), $c_i^a $ are the geminal coefficients, and the summation runs over all virtual orbitals. Specifically, the AP1roG geminal coefficient matrix has the form
    \begin{equation}\label{eq:geminal-matrix}
        \mathbf{c}_{ \rm AP1roG}=
        \begin{pmatrix}
  1      & \cdots  & 0      & 0 & c_1^{P+1} & c_1^{P+2}&\cdots &c_1^{K}\\
  0      & 1       & \cdots & 0 & c_2^{P+1} & c_2^{P+2}&\cdots &c_2^{K}\\
  \vdots & \vdots  & \ddots &   & \vdots    & \vdots   &\ddots &\vdots\\
  0      & \vdots  & \cdots & 1 & c_P^{P+1} & c_P^{P+2}&\cdots & c_P^{K}
 \end{pmatrix},
    \end{equation}
where $K$ denotes the number of basis functions, $P$ the number of electron pairs, and the left sub-block of $\mathbf{c}_{ \rm AP1roG}$ entails some reference determinant.
The geminal matrix connects each geminal with the underlying one-particle basis functions. We should note that if we impose specific restrictions on the structure of the above matrix, we can deduce different geminal models~\cite{Johnson_2013}.

The electronic wavefunction is written as a product of geminal creation operators for all electron pairs $P$ acting on the vacuum state, $\vert {\rm AP1roG} \rangle = \prod_i^P \phi_i \vert \rangle$. Unique among geminal methods, the AP1roG wavefunction ansatz can be rewritten in terms of one-particle functions as a fully general pair-Coupled-Cluster-Doubles~\cite{p-CCD} (pCCD) wavefunction, 
\begin{equation}\label{eq:ap1rog}
|{\rm AP1roG}\rangle = \exp \left (  \sum_{i=1}^P \sum_{a=P+1}^K C_i^a a_a^{\dagger}  a_{\bar{a}}^{\dagger}a_{\bar{i}} a_{i}  \right )|\Phi_0 \rangle,
\end{equation}
where $|\Phi_0 \rangle$ is some independent-particle wavefunction (for instance the Hartree--Fock (HF) determinant).
The exponential ansatz of AP1roG (\emph{cf}.~eq.~\eqref{eq:ap1rog}) ensures size-extensivity of the model. However, to recover size-consistency, we have to optimize the one-particle basis functions. This can be done in a fully variational manner~\cite{OO-AP1roG,Tamar-pCC}, analogous to orbital-optimized Coupled Cluster~\cite{Helgaker_book}, or using non-variational orbital optimization techniques~\cite{PS2-AP1roG,AP1roG-JCTC}. A number of numerical studies on systems with strongly correlated electrons showed the superiority of the variational orbital optimization procedure over the latter ones~\cite{AP1roG-JCTC}. We should note that due to the four-index transformation of the electron repulsion integrals, the computational scaling of the orbital-optimized AP1roG model deteriorates to $\mathcal{O}\big(K^5\big)$~\cite{OO-AP1roG}. Although restricted orbital-optimized AP1roG is limited to close-shell systems, it has already proven to be a reliable method for modeling strong electron correlation effects in quasi-degenerate systems~\cite{OO-AP1roG,PS2-AP1roG,AP1roG-JCTC}, single and multiple bond-breaking processes~\cite{Kasia_ijqc,pawel_jpca_2014}, and actinide chemistry~\cite{pawel_PCCP2015}. 

As all other geminal models, AP1roG misses a large fraction of weak (dynamical) electron correlation effects. To address this problem and account for weak electron correlation effects in the geminal reference wavefunction, various \textit{a posteriori} corrections have been proposed. These include models based on single- and multi-reference Perturbation Theory~\cite{Rosta2002,Rassolov2004,Piotrus_PT2,jeszenszki2014perspectives,Toth2015}, Extended Random Phase Approximation~\cite{Pernal_ERPA2014,Pastorczak2015}, (Linearized) Coupled Cluster theory~\cite{benzene-Gordon,Zoboki2013}, and Density Functional Theory~\cite{Garza2015,C5CP02773J}. In the case of AP1roG, dynamical correlation was included using Perturbation Theory~\cite{Piotrus_PT2} and Density Functional Theory~\cite{Garza2015,C5CP02773J}. Recent studies on diatomic molecules, however, point out numerical instabilities and failures of the proposed Perturbation Theory corrections~\cite{pawel_jpca_2014}. This motivates the development of different, ideally more robust dynamical correlation models for AP1roG. A reliable way to account for dynamical correlation effects \emph{a posteriori} is to use a multi-reference Linearized Coupled Cluster (LCC) correction. Recently, Zoboki \emph{et al.} presented an LCC correction based on an APSG reference function and demonstrated the good performance of the APSG-MRLCC approach. Their findings encouraged us to develop an LCC correction based on an AP1roG reference state. 

This work is organized as follows. In section~\ref{sec:lcc}, we will discuss two different LCC corrections for AP1roG. Their performance is compared by studying some well-known problems in quantum chemistry that require a balanced treatment of dynamical and strong electron correlation effects: the dissociation of C$_2$ and F$_2$, the symmetric dissociation of H$_{50}$, and spectroscopic constants of the UO$_2^{2+}$ molecule.
Computational details are presented in section~\ref{sec:comp}, while numerical results are discussed in section~\ref{sec:results}. Finally, we conclude in section~\ref{sec:conclusion}.

\section{LCC theory with an AP1roG reference function}\label{sec:lcc}
In this work, dynamical correlation effects are built in the electronic wavefunction \emph{a posteriori} using an exponential Coupled Cluster ansatz,
        \begin{equation}\label{eq:lcc}
            |\Psi \rangle = \exp({\hat{T}})  \vert {\rm AP1roG} \rangle,
        \end{equation}
where $\hat{T} = \sum_\nu t_\nu \hat{\tau}_\nu$ is a general cluster operator. The corresponding time-independent Schr\"odinger equation reads
        \begin{equation}
            \hat{H} \exp(\hat{T}) \vert {\rm AP1roG} \rangle = E \exp(\hat{T}) \vert {\rm AP1roG} \rangle
        \end{equation}

Multiplying from the left by $\exp(-\hat{T})$ and truncating the Baker--Campbell--Hausdorff expansion after the second term,
        \begin{equation}
            \exp(-\hat{T})\hat{H} \exp(\hat{T}) \approx \hat{H} + [\hat{H},\hat{T}],
        \end{equation}
we arrive at the Linearized Coupled Cluster Schr\"odinger equation
        \begin{equation}\label{eq:lccse}
            (\hat{H} + [\hat{H},\hat{T}]) \vert {\rm AP1roG} \rangle =  E \vert {\rm AP1roG} \rangle.
        \end{equation}
To obtain the cluster amplitudes $t_\nu$, we multiply from left by $\langle \nu \vert$
        \begin{equation}
            \langle \nu \vert (\hat{H} + [\hat{H},\hat{T}]) \vert {\rm AP1roG} \rangle =  0,
        \end{equation}
where we assume that the excitation operator $\hat{\tau}_\nu$ creates states orthogonal to $\vert {\rm AP1roG} \rangle$, $\langle \nu \vert {\rm AP1roG} \rangle = 0$. The projection manifold $\{\nu\}$ will depend on the choice of the cluster operator ${\hat{T}}$ (vide infra).

The energy can be calculated by projecting against the reference determinant of $\vert {\rm AP1roG} \rangle$, \emph{i.e.}, multiplying eq.~\eqref{eq:lccse} by $\langle \Phi_0 \vert$ and using intermediate normalization,
        \begin{equation}
          \hspace*{-0.5cm} \langle \Phi_0 \vert (\hat{H} + [\hat{H},\hat{T}]) \vert {\rm AP1roG} \rangle = E.
        \end{equation}

The only constraint on the cluster operator we have made so far is that it creates states that are orthogonal to the AP1roG reference function. A possible choice for the cluster operator that ensures this orthogonality condition is to include substitutions between the occupied and virtual orbitals with respect to $\vert {\rm AP1roG} \rangle$. If only double excitations are included, the cluster operator is specified as
        \begin{equation}\label{eq:t2}
            \hat{T}_2 = \frac{1}{2} \sum_{ij}^{\rm occ}\sum_{ab}^{\rm virt}{^\prime} t_{ij}^{ab} \hat{E}_{ai} \hat{E}_{bj},
        \end{equation}
where $\hat{E}_{ai} = a^\dagger_{a}a_i+a^\dagger_{\bar{a}}a_{\bar{i}}$ is the singlet excitation operator and the cluster amplitudes are symmetric with respect to pair-exchange, \emph{i.e.}, $t_{ij}^{ab}=t_{ji}^{ba}$. The prime in the above summations indicates that pair-excited determinants are excluded in the cluster operator, \emph{i.e.}, $t_{ii}^{aa} = 0$ (as those excitations do not fulfill the orthogonality condition, $\langle \nu \vert {\rm AP1roG} \rangle = 0$).

To arrive at a computationally feasible model, we will further restrict the cluster operator of eq.~\eqref{eq:t2} to allow for excitations with respect to the reference determinant only. Thereby, we exclude possible redundancies in excitations and amplitudes. The projection manifold then contains all doubly-excited determinants with respect to $\vert \Phi_0 \rangle$. Furthermore, as basis for the bra states of the projection manifold, we will use the convenient choice
        \begin{equation}
            \langle \overline{^{ab}_{ij}} \vert = \frac{1}{3} \langle ^{ab}_{ij} \vert + \frac{1}{6} \langle ^{ab}_{ji} \vert,
        \end{equation}
where $\langle ^{ab}_{ij} \vert = \langle \Phi_0 \vert \hat{E}_{jb}\hat{E}_{ia}$. The bra basis then forms a biorthogonal basis which satisfy the normalization condition
        \begin{equation}
            \langle \overline{^{ab}_{ij}} \vert ^{cd}_{kl} \rangle = \delta_{iajb,kcld} + \delta_{jbia,kcld}.
        \end{equation}
    
The doubles amplitudes $\{t_{jk}^{bc}\}$ are obtained by solving a linear set of equations
        \begin{equation}\label{eq:amplitudes}
            B_{\mu} + \sum_{\nu} A_{\mu,\nu} t_{\nu} = 0,
        \end{equation}
where the sum runs over all double excitations (without pair excitations) and $ B_{\mu} = B_{jbkc}  = \langle \overline{^{bc}_{jk}} \vert \hat{H} \vert {\rm AP1roG} \rangle$, while $ A_{\mu,\nu} = A_{jbkc,iaof} = \frac{1}{2}\langle \overline{^{bc}_{jk}} \vert [\hat{H}, \hat{E}_{ai} \hat{E}_{fo}] \vert {\rm AP1roG} \rangle$. The energy correction $E^{(D)}_{\rm corr}$ with respect to the AP1roG reference wavefunction is given as
        \begin{equation}
           E^{(D)}_{\rm corr} = \sum_{jbkc} t_{jk}^{bc} ( \langle jk \vert\vert bc \rangle + \langle jk \vert bc \rangle).
        \end{equation}
where we have used the standard notation for the exchange intergrals, $\langle jk \vert\vert bc \rangle= \langle jk \vert bc \rangle-\langle jk \vert cb \rangle$, and physicists' notation for the two-electron integrals.

Similarly, the contribution of single excitations can be accounted for by including
        \begin{equation}
            \hat{T}_1 = \sum_{ia} t_i^a \hat{E}_{ai}
        \end{equation}
in the cluster operator. Restricting the single excitations to the AP1roG reference determinant $\vert \Phi_0 \rangle$, the singles projection manifold contains all singly-excited determinants with respect to $\vert \Phi_0 \rangle$. In analogy to the doubles projection manifold, the bra states of the singles projection manifold are chosen to form a biorthogonal basis with the convenient normalization condition
        \begin{equation}
            \langle \overline{^{a}_{i}} \vert ^{b}_{j} \rangle = \delta_{ai,bj},
        \end{equation}
where $\langle \overline{^{a}_{i}} \vert = \frac{1}{2} \langle {^{a}_{i}} \vert = \frac{1}{2} \langle \Phi_0\vert \hat{E}_{ia}$. The single and double amplitudes are obtained by solving a coupled set of linear equations equivalent to eq.~\eqref{eq:amplitudes} where $\mu$ and $\nu$ now run over all single and double excitations. The energy correction with respect to the AP1roG reference value is as follows
        \begin{equation}
           E^{(S,D)}_{\rm corr} = 2\sum_{jb}F_{jb}t_j^b + \sum_{jbkc} t_{jk}^{bc} ( \langle jk \vert\vert bc \rangle + \langle jk \vert bc \rangle),
        \end{equation}
where $F_{jb}$ are the elements of the Fock matrix, $F_{jb} = h_{jb}+\sum_{m}^{\rm occ} (\langle bm \vert \vert jm\rangle+ \langle bm \vert jm\rangle)$ and $h_{jb}$ are the one-electron integrals. Note that, in contrast to canonical Hartree--Fock orbitals, the Fock matrix is not diagonal when the orbitals are optimized within AP1roG. In the AP1roG-LCC approach, the single excitations thus contribute both directly to the energy correction and indirectly through coupling to the doubles equations.

We will abbreviate the LCC correction using $\hat{T}=\hat{T}_2$ as AP1roG-LCCD, while AP1roG-LCCSD indicates that the cluster operator contains single and double excitations, $\hat{T}=\hat{T}_1+\hat{T}_2$.

\section{Computational Details}\label{sec:comp}
\subsection{AP1roG}
All geminal calculations have been performed in the \textsc{Horton 2.0.0} software package~\cite{HORTON2.0}. All restricted (variationally) orbital-optimized AP1roG calculations were allowed to freely relax without any spatial symmetry constraints, \emph{i.e.}, no point group symmetry was imposed. For all molecules, all orbitals were active. In the following, we will abbreviate (variationally) orbital-optimized AP1roG as AP1roG.

\subsection{PTa and PTb}
The PTa and PTb calculations were performed using the \textsc{Horton 2.0.0} software package~\cite{HORTON2.0}. For all molecules, the optimized AP1roG natural orbitals were taken as orbital basis and all electrons and orbitals have been included in the active space.

\subsection{LCCD and LCCSD}
The Linearized Coupled Cluster models with double and single and double excitations have been implemented in a developer version of \textsc{Horton 2.0.0}~\cite{HORTON2.0}. The natural orbitals of AP1roG were chosen as orbital basis for all molecules studied. Furthermore, all electrons and orbitals were correlated.

\subsection{Coupled Cluster}
The Coupled Cluster Doubles (CCD), CC Singles and Doubles (CCSD) as well as CC Singles, Doubles and perturbative Triples (CCSD(T)) calculations have been carried out in the DALTON2013 software package~\cite{Dalton2013}. In each case, all electrons and orbitals were correlated and no spatial symmetry was imposed.

\subsection{Relativity and Basis Sets}
For the C$_2$ and F$_2$ molecules, Dunning's aug-cc-pVDZ (C, F:(10s5p2d) $\rightarrow$ [4s3p2d]) and aug-cc-pVTZ (C, F: (11s6p3d2f) $\rightarrow$ [5s4p3d2f]) basis sets were used, while the STO-6G basis set was used for the H atoms in H$_{50}$ to allow for a comparison to DMRG reference data. 

For UO$_2^{2+}$, scalar relativistic effects were incorporated through relativistic effective core potentials (RECP). In all calculations, we have used a small core (SC) RECP (60 electrons in the core) with the following contraction scheme (12s11p10d8f) $\rightarrow$ [8s7p6d4f]. For the lighter elements (O), the cc-pVDZ basis set of Dunning was employed, (10s5p1d) $\rightarrow$ [4s3p1d].

\subsection{Fitting procedure} 
The potential energy curves of diatomic molecules were obtained by varying bond lengths in a range of $1.2-4.0$~\AA{} and $1.1-3.2$~\AA~for the F$_2$ and C$_2$ molecules, respectively.
The points on the resulting potential energy curve were used for a subsequent generalized Morse function~\cite{Coxon_1992} fit to obtain the equilibrium bond lengths (R$_{\rm e}$) and potential energy depths (D$_{\rm e}$).
The harmonic vibrational frequencies ($\omega_{\rm e}$) were calculated numerically using the five-point finite difference stencil~\cite{Abramowitz}.

\begin{table*}[t]
\begin{center}
    \caption{Spectroscopic constants for the dissociation of the C$_2$ and F$_2$ molecule for different quantum chemistry methods and basis sets. The differences are with respect to MRCI-SD reference data~\cite{Peterson1993-homo}.}\label{tab:diatomics}
    \begin{tabular}{l|l|cr@{(}lr@{(}l|cr@{(}lr@{(}l} \hline\hline
        & & \multicolumn{5}{c|}{aug-cc-pVDZ}  & \multicolumn{5}{c}{aug-cc-pVTZ} \\
        & Method & \multicolumn{1}{c}{$r_e$ [\AA{}]} & \multicolumn{2}{c}{$D_e$ [$\frac{\rm kcal}{\rm mol}$]} & \multicolumn{2}{c|}{$\omega_e$ [cm$^{-1}$]}  & \multicolumn{1}{c}{$r_e$ [\AA{}]} & \multicolumn{2}{c}{$D_e$ [$\frac{\rm kcal}{\rm mol}$]} & \multicolumn{2}{c}{$\omega_e$ [cm$^{-1}$]} \\ \hline \hline
\multirow{7}{*}{C$_2$} & AP1roG& 1.240($-$0.033) & 104.6&$-$15.7) & 1943&$+$136) & 1.227($-$0.025) & 132.9&$-$8.2)  & 1780&$-$56) \\
             &  AP1roG-PTa     & 1.273($+$0.000) & 152.7&$+$22.4) & 1902&$+$95)  & 1.251($+$0.001) & 160.4&$+$19.3) & 1889&$+$53) \\
             &  AP1roG-PTb     & 1.260($-$0.013) & 116.3&$-$14.0) & 1863&$+$56)  & 1.235($-$0.017) & 127.2&$-$13.9) & 1938&$+$102) \\
             &  AP1roG-LCCD    & 1.261($-$0.012) & 124.3&$-$7.0)  & 1926&$+$119) & 1.240($-$0.012) & 139.3&$-$1.8)  & 1916&$+$80) \\
             &  AP1roG-LCCSD   & 1.266($-$0.007) & 134.5&$+$4.2)  & 1855&$+$48)  & 1.240($-$0.012) & 143.0&$+$1.9)  & 1926&$+$90) \\
             &         NEVPT2  & 1.259($-$0.014) & 135.0&$+$4.7)  & 1924&$+$117) & 1.244($-$0.008) & 148.0&$+$6.9)  & 1886&$+$50) \\ \hline
             &         MRCI-SD & 1.273           & \multicolumn{2}{c}{130.3} & \multicolumn{2}{c|}{1807} & 1.252 & \multicolumn{2}{c}{141.1} & \multicolumn{2}{c}{1836}\\ \hline \hline

\multirow{7}{*}{F$_2$} & AP1roG      & 1.521($+$0.068) & 12.8&$-$15.7) & 886&$+$85) & 1.467($+$0.047) & 16.2&$-$7.7)  & 703&$-$189) \\
                       & AP1roG-PTa  & 1.398($-$0.055) & 30.1&$+$1.6)  & 636&$-$165)& 1.448($+$0.028) & 33.7&$-$0.2)  & 847&$-$45) \\
                       & AP1roG-PTb  & 1.473($+$0.020) & \multicolumn{2}{c}{--} & 832&$+$31) & 1.417($-$0.003) & \multicolumn{2}{c}{--} & 891&$-$1) \\
                       & AP1roG-LCCD & 1.466($+$0.013) & 39.5&$+$11.0) & 780&$-$21) & 1.433($+$0.013) & 45.5&$+$11.6) & 872&$-$20) \\
                       & AP1roG-LCCSD& 1.462($+$0.009) & 40.1&$+$11.6) & 793&$-$8)  & 1.431($+$0.011) & 46.7&$+$12.8) & 883&$-$9) \\
                       & CCSD        & 1.426($-$0.017) & 57.3&$+$18.8) & 917&$+$116)& 1.396($-$0.024) & 69.2&$+$35.3) & 1004&$+$12)\\ 
                       & CCSD(T)     & 1.450($-$0.003) & \multicolumn{2}{c}{--} & 841&$+$40) & 1.419($-$0.001) & \multicolumn{2}{c}{--} & 911&$+$19)\\ \hline
                       & MRCI-SD   & 1.453 & \multicolumn{2}{c}{28.5} & \multicolumn{2}{c|}{801} & 1.420  & \multicolumn{2}{c}{33.9} & \multicolumn{2}{c}{892} \\ \hline \hline
    \end{tabular} 
\end{center}
\end{table*}

\section{Numerical results}\label{sec:results}

\subsection{Dissociation of C$_2$}

\begin{figure}[b]
\includegraphics[width=0.45\textwidth]{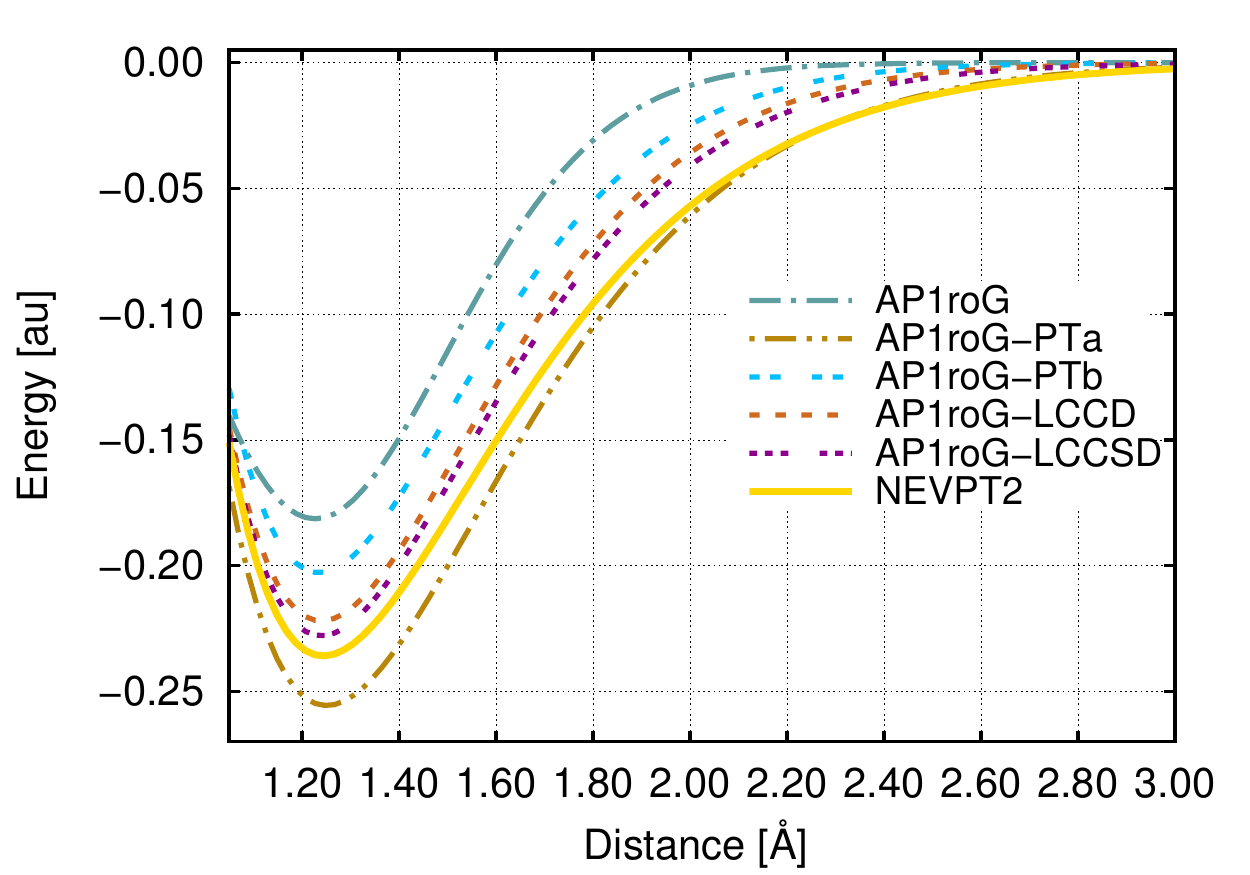}
\caption{Potential energy surfaces for the dissociation of the C$_2$ molecule using an aug-cc-pVTZ basis set.}\label{fig:c2}
\end{figure}

The carbon dimer is one of the most complex diatomic molecules that can be formed from the first-row elements of the periodic table. The unusual nature of the carbon-carbon bond and the question concerning its bond order attracted a lot of attention from theoretical chemists~\cite{Shaik2012,Matxain2013,Markus-chemistry,PCCP_bonding} in recent years. 
Around the equilibrium distance, the electronic structure of the C$_2$ molecule has two dominant configurations $1\sigma_g^21\sigma_u^2 2\sigma_g^2 2\sigma_u^2 1 \pi_u^4$ and $1\sigma_g^21\sigma_u^2 2\sigma_g^2 1 \pi_u^4 3\sigma_g^2$ as well as a number of other configurations arising from low-lying excited states~\cite{Peterson1993-homo,Sherill_C2,PHF_excited-states,CheMPS2}. When the two carbon atoms are pulled apart, the molecular system becomes strongly multi-reference. However, even for stretched carbon-carbon distances, dynamical electron correlation effects remain non-negligible~\cite{Sherill_C2,CheMPS2,pawel_jpca_2014}. 
A reliable theoretical description of spectroscopic constants (bond lengths, potential energy well depths, and vibrational frequencies) thus requires a balanced treatment of all types of electron correlation effects~\cite{entanglement_letter,Kasia_ijqc} (static, non-dynamic, and dynamic) along the dissociation pathway.
Since highly accurate reference data for the spectroscopic constants of C$_2$ is available, it is an ideal candidate to test our AP1roG-LCC approach. 

The upper part of Table~\ref{tab:diatomics} summarizes the spectroscopic constants of the C$_2$ molecule for different basis sets and quantum chemistry methods including various dynamical correlation models based on an AP1roG reference function. As expected, AP1roG predicts too short equilibrium bond distances and too shallow potential well depths for all basis sets studied. If dynamical correlation is included \emph{a posteriori} on top of the AP1roG reference function using perturbation theory, spectroscopic constants improve only slightly compared to MRCI-SD reference data. Although AP1roG-PTa predicts equilibrium bond distances that are in perfect agreement with MRCI-SD, the potential well depth is overestimated and the differences with respect to MRCI-SD are even larger than for AP1roG without PTa correction. Furthermore, AP1roG-PTb does not improve potential well depths and vibrational frequencies compared to AP1roG when the basis set is enlarged from aug-cc-pVDZ to aug-cc-pVTZ. In contrast to the PTa and PTb models, an LCC correction on top of AP1roG including doubles and singles and doubles yields spectroscopic constants that are in very good agreement with MRCI-SD reference data, outperforming NEVPT2 (differences are less than 2 kcal/mol for potential well depths using aug-cc-pVTZ).

Figure~\ref{fig:c2} shows the fitted potential energy surfaces for the aug-cc-pVTZ basis set. We should note that all potential energy surfaces were adjusted to Zero in the dissociation limit. Comparing the spectroscopic constants of Table~\ref{tab:diatomics}, we can conclude that the MRCI-SD potential energy surface would lie between the AP1roG-LCCD and AP1roG-LCCSD potential energy curves. All other quantum chemistry methods yield potential energy surfaces that deviate more from the expected MRCI-SD reference curve.

\subsection{Dissociation of F$_2$}

F$_2$ is a well-known example of diatomic molecules where dynamical electron correlation effects play a dominant role~\cite{PCCP_bonding}. 
Furthermore, theoretical studies indicate that large basis sets and robust dynamical electron correlation models are required to reproduce the experimentally determined 
spectroscopic constants~\cite{Jankowski_F2,Ahlrichs_bs,Paldus_F2,Piecuch_F2,Ivanov_F2,Monika_F2}. These features make the F$_2$ molecule a good test case to assess the reliability of the LCC correction on top of an AP1roG reference function.     

\begin{figure}[t]
\includegraphics[width=0.45\textwidth]{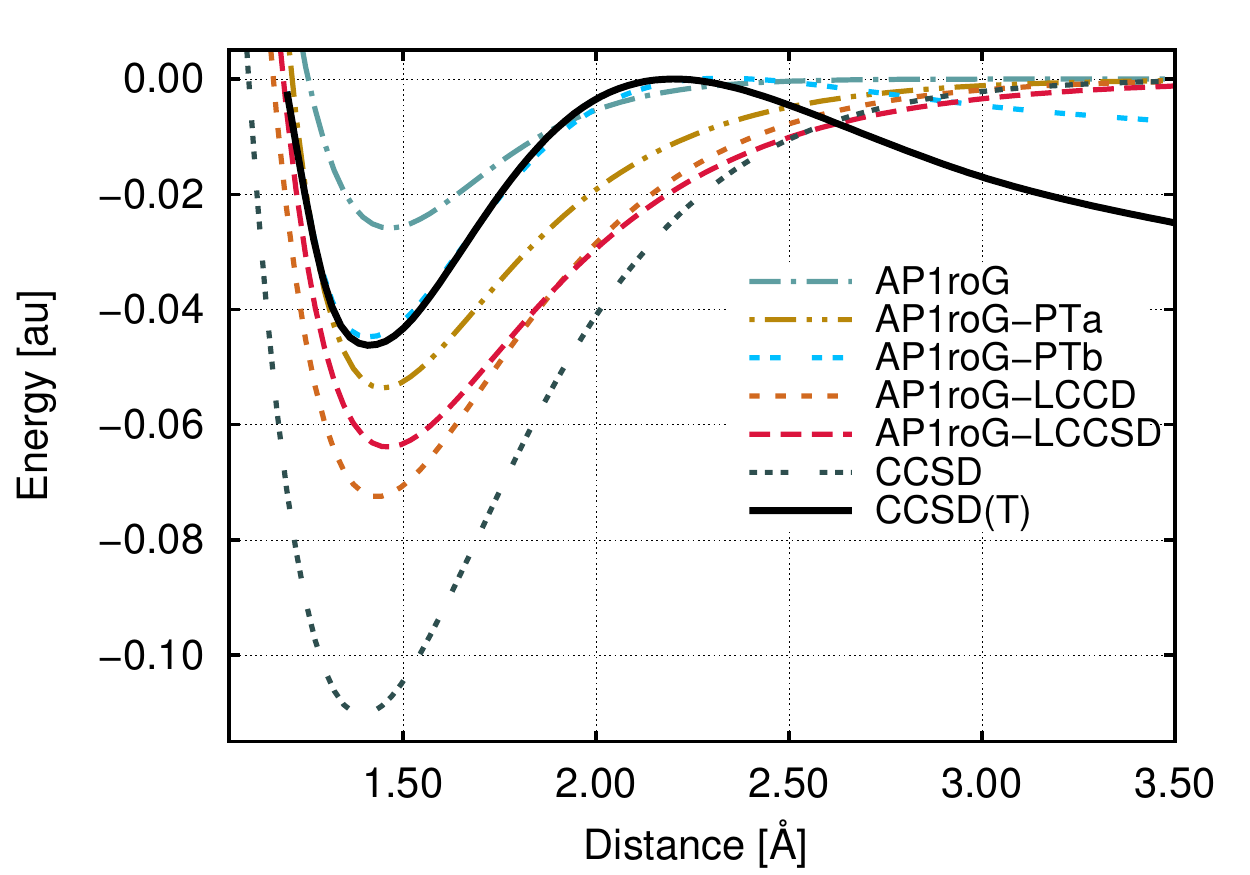}
\caption{Potential energy surfaces for the dissociation of the F$_2$ molecule using a aug-cc-pVTZ basis set.}\label{fig:f2}
\end{figure}

The bottom part of Table~\ref{tab:diatomics} summarizes the spectroscopic constants for the dissociation process of the F$_2$ molecule using different basis sets and dynamical correlation models. We should note that both CCSD(T) and AP1roG-PTb diverge in the dissociation limit and thus the potential energy well depths are not given in Table~\ref{tab:diatomics} (see also Figure~\ref{fig:f2}). To obtain an estimate for $D_e$, we have taken the MRCI-SD results by Peterson~\cite{Peterson1993-homo}. Note that $r_e$ and $\omega_e$ as predicted by MRCI-SD are in good agreement with CCSD(T) calculations.

As observed for the C$_2$ molecule, AP1roG considerably overestimates the equilibrium bond length and underestimates the potential energy depth, which can be attributed to the large fraction of dynamical correlation that cannot be captured by restricting the wavefunction to electron-pair states. Although AP1roG-PTa yields potential energy depths that are in very good agreement with MRCI-SD reference data, equilibrium bond lengths and vibrational frequencies deviate considerably from the MRCI-SD reference values. Specifically, when increasing the basis set from aug-cc-pVDZ to aug-cc-pVTZ, the predicted equilibrium bond length changes from being too short to being overestimated. In contrast to PTa, AP1roG-PTb results in equilibrium bond lengths and vibrational frequencies that are in perfect agreement with MRCI-SD reference data when the basis set is increased to triple-zeta quality, fails however in the vicinity of dissociation. The LCC correction on top of AP1roG results in the most stable and robust dynamical correlation model, yielding similar results for increasing basis set sizes and outperforming CCSD. Specifically, equilibrium bond lengths and vibrational frequencies are in very good agreement with MRCI-SD reference data. However, potential energy well depths are considerably overestimated by AP1roG-LCC (around 12 kcal/mol with respect to MRCI-SD), which might be attributed to restricting the cluster operator to singles and doubles excitations~\cite{Jankowski_F2}(\emph{cf.}, CCSD overestimates $D_e$ by more than 20-35 kcal/mol, depending on the basis set size).

\subsection{Symmetric Dissociation of H$_{50}$}
The symmetric stretching of hydrogen chains is commonly used as a molecular model for strongly correlated systems and remains a challenging problem for conventional quantum-chemistry methods~\cite{Hachmann_H50,Gustavo_H50,Stella_H50,DMFT_H50,OO-AP1roG,PS2-AP1roG}. Recently, we have shown that AP1roG accurately describes the potential energy surface of the symmetric dissociation of H$_{50}$ in the vicinity of dissociation, but deviates from DMRG reference data around the equilibrium and for stretched interatomic distances~\cite{OO-AP1roG}, which can be attributed to the missing dynamical correlation energy.

Table~\ref{tab:h50} summarizes the non-parallelity error (NPE) per hydrogen atom for the symmetric dissociation of H$_{50}$ obtained by AP1roG and different dynamical correlation models. The large NPE per hydrogen atom of AP1roG can be associated with the missing dynamical correlation energy around the equilibrium distance and for stretched interatomic bond lengths. Adding dynamical correlation \emph{a posteriori} improves the NPE per hydrogen atom considerably. While PTa, PTb, and LCCD have a similar NPE per hydrogen atom of about 4.5 m$E_h$, the NPE is reduced to less than 1.5 m$E_h$ if single excitations are included in the cluster operator.

\begin{figure}[t]
\includegraphics[width=0.45\textwidth]{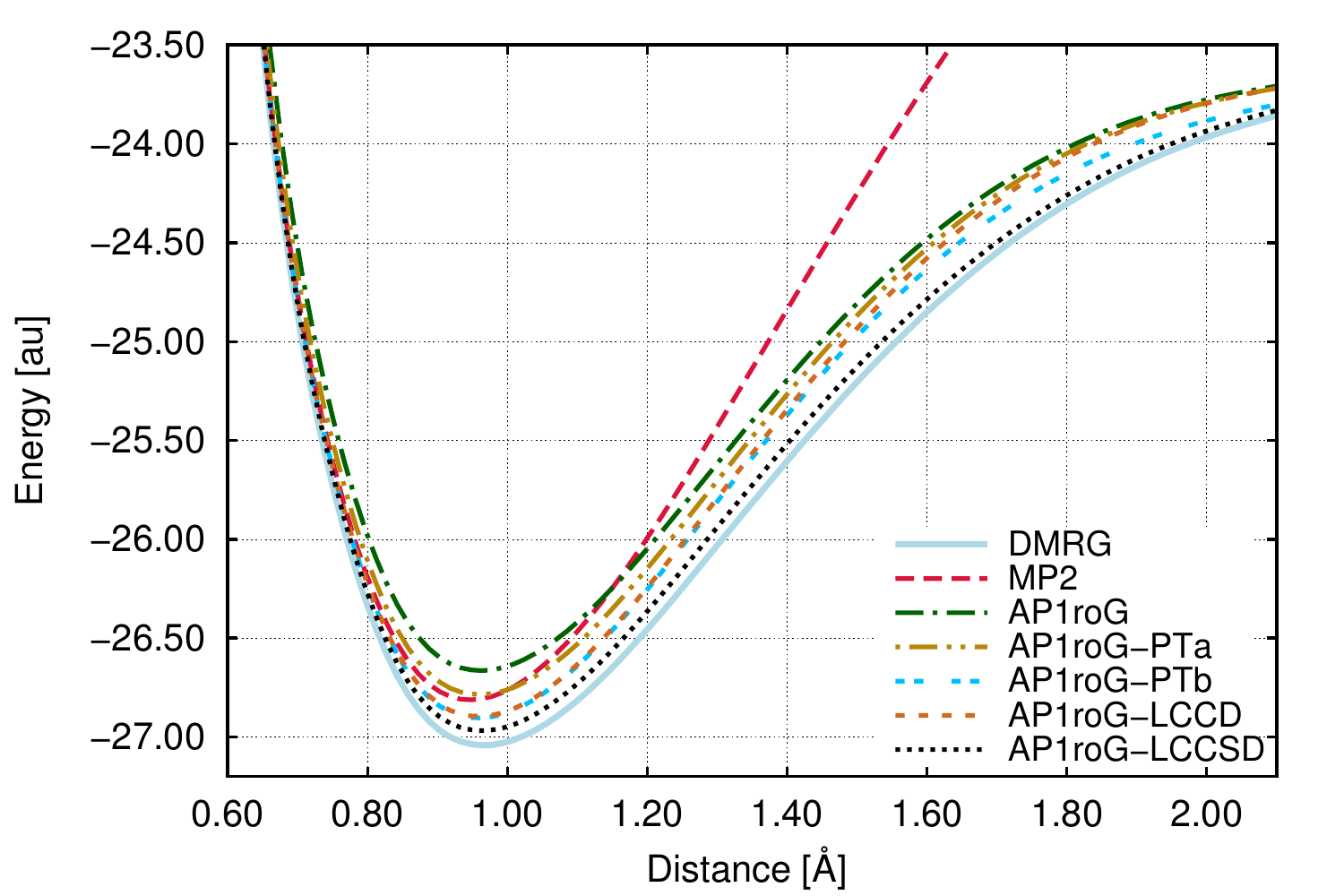}
\caption{Symmetric dissociation of the H$_{50}$ chain using the STO-6G basis set obtained from different methods. The DMRG reference data are taken from Ref.~\cite{Hachmann_H50}, while the MP2 and AP1roG data are taken from Ref.~\cite{OO-AP1roG}.}\label{fig:h50}. 
\end{figure}

The importance of single excitations in the LCC model is also noticeable in the shape of the potential energy surface shown in Figure~\ref{fig:h50}. While AP1roG-PTa, AP1roG-PTb, and AP1roG-LCCD have similar potential energy curves (in terms of shape and total electronic energies) with AP1roG-PTb being lowest in energy, the potential energy surface obtained by AP1roG-LCCSD considerably deviates from the aforementioned dynamical correlation models for short and intermediate bond lengths (around 1.0 and 1.5 \AA{}). Furthermore, the AP1roG-LCCSD potential energy curve is in very good agreement with DMRG reference data (see Figure~\ref{fig:h50}).

\begin{table}[t]
\begin{center}
\caption{Non-parallelity error per hydrogen atom for the symmetric dissociation of the H$_{50}$ chain with respect to DMRG reference data. The MP2 and AP1roG data are taken from Ref.~\cite{OO-AP1roG}.}\label{tab:h50}
    \begin{tabular}{l|r@{.}l} \hline\hline
        Method      &  \multicolumn{2}{c}{NPE/H [m$E_h$]} \\ \hline
        MP2         & 43&540 \\  
        AP1roG      &  6&187 \\   
        AP1roG-PTa  &  4&831 \\   
        AP1roG-PTb  &  4&170 \\    
        AP1roG-LCCD &  4&506 \\    
        AP1roG-LCCSD&  1&389 \\
        \hline \hline
    \end{tabular} 
\end{center}
\end{table}

\subsection{Symmetric dissociation of UO$_2^{2+}$}

The uranyl cation (UO$_2^{2+}$) is a small building block of a large variety of uranium-containing complexes~\cite{Hayton2013,denning07,Gomes_Cs2UO2Cl4}. This molecule has a linear structure and a singlet ground-state electronic configuration. Its characteristic symmetric and asymmetric U--O vibrational frequencies are used to identify the presence of UO$_2^{2+}$ in larger molecular assemblies~\cite{denning07,Vallet_2012,Tecmer2012}. While the electronic structure of the uranyl cation is well-known around the equilibrium structure~\cite{denning_91b,uranyl_de_Jong_99,matsika_2001,real07,pierloot05,Jackson2008,real09,pawel1,pawel_saldien,pawel5,pawel_PCCP2015}, the complicated nature of the U--O bond hampers a theoretical description at larger U--O distances using standard quantum chemistry approaches~\cite{pawel_PCCP2015}. One of the limiting factors that impede theoretical studies is the large number of strongly-correlated electrons distributed among 5$f$-, 6$d$-, and 7$s$-orbitals. In addition, the 6$s$- and 6$p$-core-valence orbitals are easily polarizable and have a non-negligible contribution to the correlation energy. However, around the equilibrium structure, the uranyl cation is well described by single-reference CC theory if all important electrons are correlated. This allows us to assess the performance of the LCC models in describing dynamical correlation effects ordinating from the 5$f$-, 6$d$-, and 7$s$- as well as the core-valence electrons.

\begin{table}[t]
\begin{center}
\caption{Spectroscopic constants for the symmetric dissociation of the UO$_2^{2+}$ molecule for different quantum chemistry. The differences are with respect to CCSD(T) reference data. The CASSCF and CC data were taken from Ref.~\citenum{pawel_PCCP2015}.}\label{tab:uo2}
    \begin{tabular}{lr@{(}lr@{(}l} \hline\hline
        Method & \multicolumn{2}{c}{$r_e$ [\AA{}]} & \multicolumn{2}{c}{$\omega_e$ [cm$^{-1}$]} \\ \hline \hline
        AP1roG      & 1.669&$-$0.047) & 1062&$+$53) \\
        AP1roG-PTb  & 1.715&$-$0.001) & 1340&$+$331) \\
        AP1roG-LCCD & 1.708&$-$0.008) & 1073&$+$64) \\
       CAS(10,10)SCF& 1.694&$-$0.022) & 1079&$+$70)\\ 
       CAS(12,12)SCF& 1.707&$-$0.009) & 1034&$+$25)\\ 
        CCD         & 1.690&$-$0.026) & 1125&$+$116)\\ 
        CCSD        & 1.697&$-$0.019) & 1068&$+$59)\\ \hline
CCSD(T)&\multicolumn{2}{c}{1.716} & \multicolumn{2}{c}{1009} \\ \hline\hline
    \end{tabular} 
\end{center}
\end{table}

\begin{figure}[t]
\includegraphics[width=0.45\textwidth]{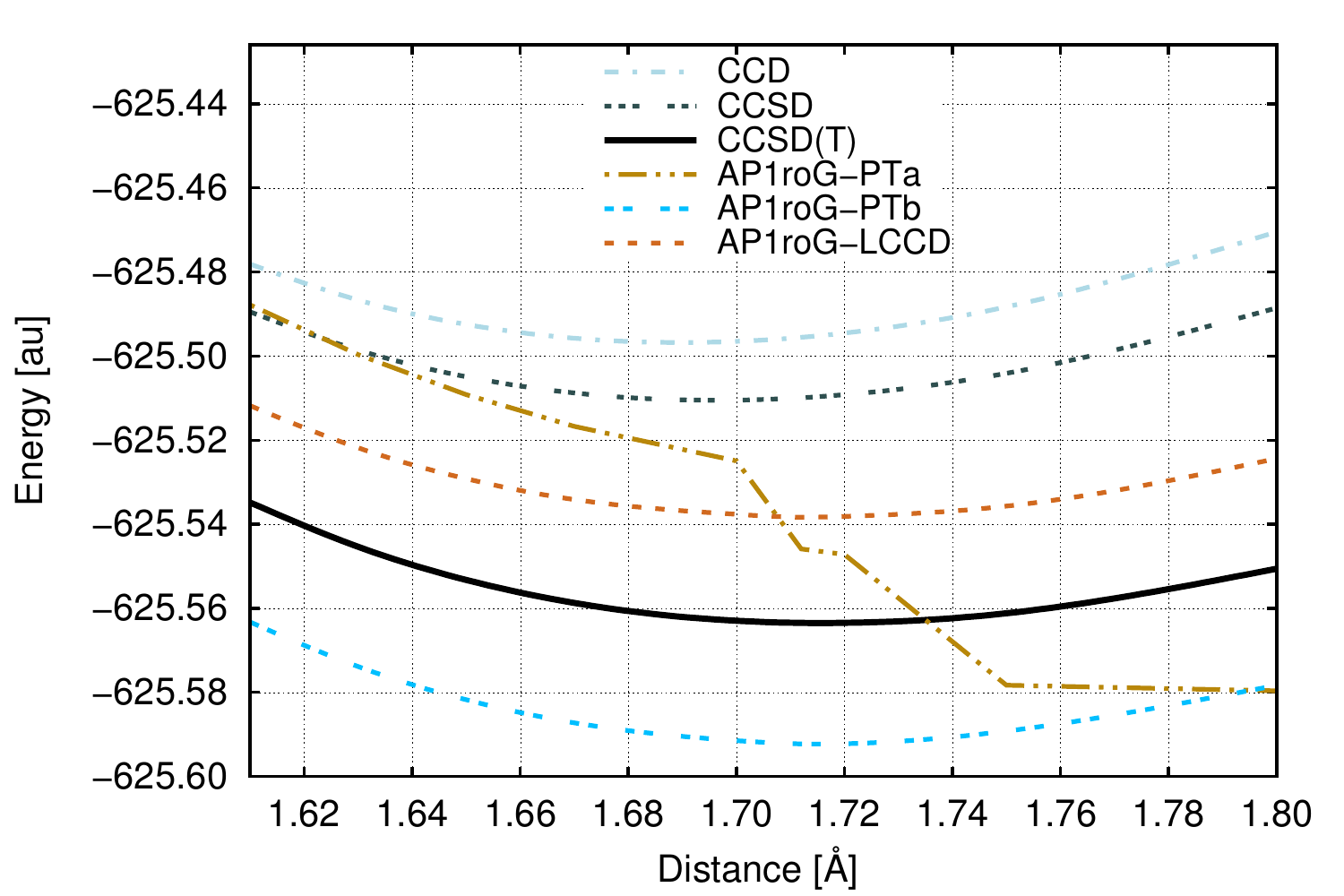}
\caption{Potential energy surfaces for the symmetric stretching of the UO$_2^{2+}$ molecule around the equilibrium geometry. Note that the CASSCF potential energy surfaces are much higher in energy and are thus not shown.}\label{fig:uo2}
\end{figure}

The equilibrium bond lengths and vibrational frequencies obtained by different quantum chemistry methods are shown in Table~\ref{tab:uo2}. As expected, AP1roG considerably underestimates the equilibrium bond length, while $\omega_e$ is in good agreement with CCSD(T). Adding dynamical correlation effects on top of AP1roG shifts $r_e$ closer to CCSD(T) reference data. The shape of the potential, however, strongly depends on the AP1roG dynamical correlation model. Specifically, PTb results in a much steeper potential energy surface overestimating vibrational frequencies by more than 330 cm$^{-1}$ compared to CCSD(T), while LCCD preserves the shape of the potential energy surface and yields a vibrational frequency that agrees well with AP1roG and CCSD(T) data (differences amount to ca.~60 cm$^{-1}$). The overall accuracy of AP1roG-LCCD lies between CCSD and CCSD(T), being closer to the latter. We should emphasize that PTa completely fails for the UO$_2^{2+}$ molecule and produces a discontinuous potential energy surface around the equilibrium (see also Figure~\ref{fig:uo2}). Furthermore, the CASSCF equilibrium distance strongly depends on the size of the active space chosen in CASSCF calculations. Specifically, increasing the active space from CAS(10,10) to CAS(12,12), \emph{i.e.}, including the $\sigma$- and $\sigma^*$-orbitals, results in spectroscopic constants that are in good agreement with AP1roG-LCCD and CCSD(T) data.

Figure~\ref{fig:uo2} shows the fitted potential energy surfaces around the equilibrium for selected quantum chemistry methods. AP1roG-LCCD yields total electronic energies that are between CCSD and CCSD(T), while the potential energy surface predicted by AP1roG-PTb is considerably lower than the CCSD(T) reference curve. Note that the potential energy surfaces optimized by CASSCF lie much higher in energy and are thus not shown in Figure~\ref{fig:uo2}.

\section{Conclusions}\label{sec:conclusion}

We have presented an alternative model to capture dynamical correlation effects on top of an AP1roG reference functions that uses a Linearized Coupled Cluster ansatz. Our approach is motivated by previous studies of an LCC correction to an APSG reference function~\cite{Zoboki2013}. Specifically, our cluster operator is restricted to doubles and singles and doubles excitations as in the standard Coupled Cluster approach, \emph{i.e.}, excitations from the occupied to the virtual orbitals of some reference determinant.
We have compared this new dynamical correlation ansatz to the PTa and PTb perturbation theory models as well as standard quantum chemistry approaches for the C$_2$ and F$_2$ molecules, the H$_{50}$ hydrogen chain, and the uranyl cation, UO$_2^{2+}$.

In general, both AP1roG-LCC models yield similar spectroscopic constants and are closest to MRCI-SD, CCSD(T), and DMRG reference data for all molecules we have investigated. Furthermore, LCCD and LCCSD represent more robust and reliable dynamical correlation models than PTa and PTb. Our study demonstrates that the success and failures of PTa and PTb are difficult to anticipate {a priori} and are strongly system-dependent. While PTa yields reasonable results for equilibrium bond lengths of the C$_2$ molecule and for the potential energy depth of the F$_2$ molecule, the corresponding C$_2$ potential energy depth and the F$_2$ equilibrium bond length significantly differ from reference data. A similar behavior was observed for PTb. In contrast to the perturbation theory models, the LCC ansatz is able to capture different flavours of dynamical correlation effects reliably, as present in the C$_2$, F$_2$, H$_{50}$, and UO$_2^{2+}$ molecules.


\section{Acknowledgment}

We gratefully acknowledge financial support from the Natural Sciences and Engineering Research Council of Canada. K.B.~acknowledges the financial support from the Swiss National Science Foundation (P2EZP2 148650), the Banting Postdoctoral Fellowship program, and the National Science Center (Grant No. DEC-2013/11/B/ST4/00771).
We had many helpful discussions with Pawe\l{} Tecmer.

\providecommand*{\mcitethebibliography}{\thebibliography}
\csname @ifundefined\endcsname{endmcitethebibliography}
{\let\endmcitethebibliography\endthebibliography}{}

\end{document}